\documentstyle[multicol,aps,epsf,epsfig]{revtex}
\begin{document}
\draft
\def\l{\lambda}
\def\e{\epsilon}
\def\d{\delta}
\def\half{{1\over2}}
\def\O{{\cal O}}
\def\qc{{q_{\rm c}}}
\def\lc{{\l_{\rm c}}}
\def\av#1{\langle#1\rangle}
\def\a{\alpha}
\def\etal{{\it et al.}}
\def\pc{p_{\rm c}}
\def\df{d_{\rm f}}
\def\K{{\tilde K}}
\def\p{{\tilde p}}
\def\P{{\tilde P}}
\title {Percolation in Directed Scale-Free Networks}
\author{N.~Schwartz$^{1}$, R.~Cohen$^{1}$, D.~ben-Avraham$^{2}$, 
A.-L.~Barab\'asi$^{3}$ and S.~Havlin$^{1}$}
\address{$^1$Minerva Center and Department of Physics, Bar-Ilan
University, Ramat-Gan, Israel}
\address{$^2$Department of Physics, Clarkson University, Potsdam NY 13699-5820,
USA}
\address{$^3$Department of Physics, University of Notre Dame, Notre Dame IN
46556, USA}
\maketitle
\begin{abstract}
Many complex networks in nature have directed links, a property
that affects the network's navigability and large-scale topology.
Here we study the percolation properties of such directed
scale-free networks with correlated {\it in} and {\it out} degree
distributions. We derive a phase diagram that indicates the
existence of three regimes, determined by the
values of the degree exponents.
In the first regime we regain the known directed percolation mean field 
exponents.
In contrast, the second and third regimes are characterized by anomalous
exponents, which we calculate analytically.
In the third regime the network is resilient to random dilution, i.e., the 
percolation threshold is $p_{c} \to 1$. 
\end{abstract}
\begin{multicols}{2}

Recently the topological properties of large complex networks
such as the Internet, WWW, electric power grid, cellular and
social networks have drawn considerable attention ~\cite{bar_rev,dor_rev}. 
Some of these networks are directed, for example,
in social and economical networks if node $A$ gains information or acquires 
physical goods from node $B$, it does not necessarily mean
that node $B$ gets similar input from node $A$. Likewise, most metabolic 
reactions are one-directional, thus changes in the concentration of molecule 
$A$ affect the concentration of its product $B$, but the reverse is not true. 
Despite the directedness of many real networks, the modeling literature,
with few notable exceptions \cite{newman,dor_samu,sanchez}, has 
focused mainly on undirected networks.

An important property of directed networks can be captured by studying their 
degree distribution, ${P(j,k)}$, or the probability that an
arbitrary node has $j$ incoming and $k$ outgoing edges. Many naturally
occurring directed networks, such as the WWW, metabolic networks, 
citation networks, etc., exhibit a
power-law, or {\it scale-free\/} degree distribution for the
incoming or outgoing links:
\begin{equation}
\label{Pk}
P_{in(out)}(l)=c l^{-\lambda_{in(out)}},\quad l\geq m\;,
\end{equation}
where $m$ is the minimal connectivity (usually taken to be $m=1$),
 $c$ is a normalization factor and $\lambda_{in(out)}$ are the in(out) degree
 exponents characterizing the network ~\cite{bar_degree,broder_degree}.
An important property of scale-free networks is their robustness to 
random failures, coupled with an increased vulnerability to 
attacks~\cite{bar2,cohen,cal,cohen2,sole}.
Recently it has been recognized that this feature can be addressed analytically
in quantitative terms \cite{cohen,cal,cohen2} by combining graph theoretical
concepts with ideas from percolation theory~\cite{stauffer,havlin,hinrichsen}. 
Yet, while the percolative properties of undirected networks
are much studied, little is known about the effect of node failure
in directed networks. As many important networks are directed, 
it is important to fully understand the
implications to their stability.
Here we show that directedness has a strong impact on
the percolation properties of complex networks and we draw a detailed phase
diagram.

\begin{figure}
\narrowtext
\hskip 0.3 in 
\includegraphics[width=0.3\textwidth]{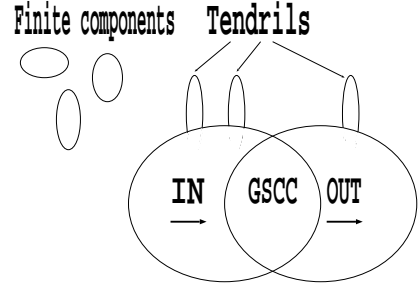} 
\hskip 0.1in
\caption{Structure of a general directed graph.} 
\label{fig1}
\end{figure}

The structure of a directed graph has been characterized in 
\cite{newman,dor_samu}, and in the context of the WWW in \cite{broder_degree}.
In general, a directed graph consists of a giant weakly connected component
(GWCC) and several finite components. In the GWCC every site is reachable from 
every other, provided that the 
links are treated as bi-directional. The GWCC is further divided into a giant
strongly connected component (GSCC), consisting of all sites reachable from 
each other following directed links. All the sites 
reachable from the GSCC are referred to as the giant OUT component, and the 
sites from which the GSCC is reachable are referred to as the giant IN 
component. The GSCC is the intersection of the IN and OUT components.
All sites in the GWCC, but not in the IN and OUT components are 
referred to as the ``tendrils'' (see Fig. \ref{fig1}.).

\noindent

For a directed random network of arbitrary degree distribution
the condition for the existence of a giant component can be deduced in a 
manner similar to \cite{cohen}. If a site, $b$, is
reached following a link pointing to it from site $a$, then it must have at 
least one 
outgoing link, on average, in order to be part of a giant component. 
This condition can be written as
\begin{equation}
\label {criterion}
\av{k_b|a\rightarrow b}=\sum_{j_b,k_b}k_b P(j_b,k_b|a\rightarrow b)=1,
\end {equation} 
where $j$ and $k$ are the {\it in-} and {\it out-} degrees, respectively, 
$P(j_b,k_b|a\rightarrow b)$ is the conditional probability given that
site $a$ has a link leading to $b$, and $\av{k_b|a\rightarrow b}$ is the
conditional average. Using Bayes rule we get 
\begin{eqnarray*}
P(j_b,k_b|a\rightarrow b)&=&P(j_b,k_b,a\rightarrow b)/P(a\rightarrow b)\\
&=&P(a\rightarrow b|j_b,k_b)P(j_b,k_b)/P(a\rightarrow b).
\end{eqnarray*}
For random networks  
$P(a\rightarrow b)=\av{k}/(N-1)$ and
 $P(a\rightarrow b|j_b,k_b)=j_b/(N-1)$, where $N$ is the total number of nodes 
in the network. The above criterion thus reduces to \cite{newman,dor_samu}
\begin{equation}
\label{jk_av}
\label{condition_general}
\av{jk}\geq\av{k}.
\end {equation}

Suppose a fraction $p$ of the nodes is removed from the
network.  (Alternatively, a fraction $q=1-p$ of the nodes is retained.)
The original degree distribution, $P(j,k)$, becomes
\begin{eqnarray}
\label{new_dist1}
&&P'(j,k)=\sum_{j_0,k_0}^\infty P(j_0,k_0)\Big({j_0\atop
j}\Big)(1-p)^{j} p^{j_0-j}  \nonumber\\
&&\ \ \times\Big({k_0\atop k}\Big)(1-p)^{k} p^{k_0-k}\;.
\end{eqnarray}
In view of this new distribution, Eq.~(\ref{jk_av}) yields the percolation
threshold
\begin{equation}
\label{gen_perc}
q_{c}=1-p_{c}= {\av{k}\over\av{jk}}\;,
\end {equation}
where averages are computed with respect to the original
distribution before dilution, $P(j,k)$.
Eq.~(\ref{gen_perc}) indicates that in directed scale-free networks 
if $\av{jk}$ diverges
then $q_{c} \rightarrow 0$ and the network is resilient to random breakdown of
nodes and bonds.

The term $\av{jk}$ may be
dramatically influenced by the appearance of correlations between the
{\it in}- and {\it out}-degrees of the nodes. In
particular, let us consider scale-free distributions for both the
{\it in}- and {\it out}-degrees:
\begin{equation}
\label{pin}
P_{in}(j)\sim\cases{
{Bc_{in} j^{-\lambda_{in}}}
&${j\not=0},$\cr {1-B} &${j=0},$ }
\end{equation}
and
\begin{equation}
P_{out}(k)=c_{out} k^{-\lambda_{out}}\;.
\end{equation}
In~(\ref{pin}) we choose to add the possible zero value to the {\it in}-degree
in order to maintain
$\av{j}=\av{k}$. If the {\it in}- and {\it out}-degrees are 
uncorrelated, we expect $\av{jk}=\av{j}\av{k}$. For several real
directed networks this equality does not hold. For example, 
the network of Notre-Dame University WWW \cite{bar_degree}, 
has $\av{k}=\av{j}\approx 4.6$, and thus $\av{j}\av{k}=21.16$. 
In contrast, measuring directly 
we find $\av{jk}\approx 200$, about an order of magnitude larger
than the result expected for the uncorrelated case. This yields an estimate of
$q_{c}\approx 0.02$, i.e., a very stable directed network. We obtained similar
results also for some metabolic networks \cite{bar3}, indicating that
in real directed networks, the {\it in}- and {\it out}-degrees are correlated.

\begin{figure}
\narrowtext
\hskip 0.3 in
\vskip 0.02in 
\includegraphics[width=0.3\textwidth]{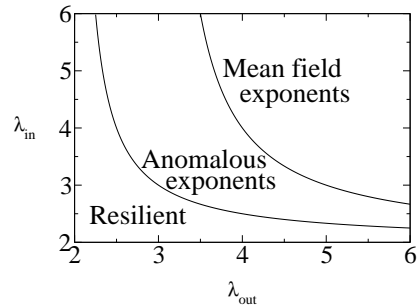}
\vskip 0.1in 
\caption{Phase diagram of
the different regimes for the IN component of scale-free correlated directed 
networks.
The boundary between Resilient and Anomalous exponents is derived from
Eq.~(\ref{boundary1}) while that between Anomalous exponents and Mean field
exponents  is given by Eq.~(\ref{boundary2}) for $\l^{\star}=4$.
For the diagram of the OUT component $\l_{in}$ and $\l_{out}$ change roles.} 
\label{fig2}
\end{figure}

To address correlations, we model it in the following manner: 
we first generate the $j$ values for the entire network. Next, for each
site with $j\not=0$ with probability $A$ we generate $k$ fully
correlated with $j$, i.e., $k=k(j)$. Assuming that $k(j)$ is a 
monotonically increasing
function then the requirement $c_{out}k^{-\l_{out}}dk=c_{in}j^{-\l_{in}}dj$ ---
needed to maintain the distributions scale-free --- leads to
$k^{\l_{out}-1}=j^{\l_{in}-1}$. With probability $1-A$, the degree $k$ is
chosen independently from $j$:
\begin{equation}
\label{pjk}
P(j,k)\sim\cases{
{(1-A)Bc_{in}j^{-\l_{in}}c_{out}k^{-\l_{out}}}
&${}$\cr {+BAc_{out}k^{-\l_{out}}\delta_{j,j(k)} } &${j\not=0},$
\cr {(1-B)c_{out}k^{-\l_{out}}} &${j=0},$}
\end{equation}
where $j(k)=k^{{\l_{out}-1\over\l_{in}-1}}$. 
With this distribution, any finite fraction $BA$ of fully
correlated sites yields a diverging $\av{jk}$ whenever
\begin{equation}
\label{boundary1}
(\l_{out}-2)(\l_{in}-2)\leq 1\;,
\end{equation}
causing the percolation threshold to vanish (see Fig. \ref{fig2}.). 

\noindent

In the case of no correlations between the {\it in}- and the
{\it out}-degrees, $A=0$, Eq.~(\ref{pjk}) becomes
${P(j,k)}=P_{in}(j)P_{out}(k)$.
Then the condition for the existence of a giant component is: 
$\av{k}=\av{j}=1.$
Moreover, Eq.~(\ref{gen_perc}) reduces to:
\begin{equation}
\label{perc}
q_{c}=1-p_{c}= {1\over\av{k}}\;.
\end {equation}
Applying~(\ref{perc}) to scale-free networks one
concludes that for $\lambda_{out}>2$ and $\lambda_{in}>2$ a phase 
transition exists at a
finite $q_{c}$. Here we concern
ourselves with the critical exponents associated with the percolation
transition in scale-free network of $\lambda_{out}>2$ and $\lambda_{in}>2$ 
which is the most relevant regime (Fig. \ref{fig2}). 

Percolation of the GWCC can be seen to be similar to percolation in
the non-directed graph created from the directed graph by ignoring the
directionality of the links. The threshold is obtained from the criterion
\cite{cohen}
\begin{equation}
\qc={\av{k}\over\av{k(k-1)}}.
\end{equation}
Here the connectivity distribution is the convolution of
the {\it in} and {\it out} distributions
\begin{equation}
P'(k)=\sum_{l=0}^k P(l,k-l).
\end{equation}

Regardless of correlations, $P'(k)$ is always dominated by the slower 
decay-exponent, 
therefore percolation of the GWCC is the same as in non-directed 
scale-free
networks, with $\l_{eff}=min(\l_{in},\l_{out})$.  Note that the percolation
threshold 
of the GWCC may differ from that of the GSCC and the IN and OUT
components~\cite{dor_samu}.

We now use the formalism of
generating functions~\cite{wilf,weiss} to analyze percolation of the 
GSCC and IN and OUT components.  
In~\cite{newman,dor_samu} a generating function is built for the joint 
probability
distribution of outgoing and incoming degrees, before dilution:
\begin{equation}
\label{PHI}
\Phi(x,y)=\sum_{k,j}P(j,k)x^jy^k\;.
\end {equation}
Using the approach of Callaway $\sl{et} \ \sl{al}$~\cite{cal} let $q(j,k)$ be 
the
probability that a vertex of degree $(j,k)$ remains in the network following
dilution.  The generating function after
dilution is then
\begin{equation}
\label{PSI}
G(x,y)=\sum_{k,j}P(j,k)q(j,k)x^j y^k\;.
\end {equation}
From~(\ref{PSI}) it is possible to define the generating function for
the outgoing degrees $G_{0}$
\begin{equation}
\label{G0}
G_{0}(y)\equiv G(1,y)=\sum_{k,j}P(j,k)q(j,k)y^k\;.
\end {equation}
The probability of reaching a site by following a specific link is
proportional to $jP(j,k)$, therefore, the probability to reach an
occupied site following a specific directed link is generated by
\begin{equation}
\label{G11}
G_{1}(y)={{\sum_{j,k} jP(j,k)q(j,k)y^k}\over{\sum_{j,k}jP(j,k)}}\;.
\end {equation}

Let $H_1(y)$ be the generating function for the probability
of reaching an outgoing component of a given size by following a directed
link, after a dilution. $H_1(y)$ satisfies the
self-consistent equation:
\begin{equation}
\label{H11}
H_{1} (y)=1-G_{1}(1)+y G_{1} (H_{1} (y))\;.
\end {equation}
Since $G_{0}(y)$ is the generating function for the outgoing degree of a site,
the generating function for the probability that $n$ sites are 
reachable from a given site is
\begin{equation}
\label{H01}
H_0 (y)=1-G_{0}(1)+yG_{0} (H_{1} (y))\;.
\end {equation}
For the case where correlations exist, and assuming random dilution: 
$q(j,k)=q$,
Eqs.~(\ref{H11}) and~(\ref{H01}) reduce to
\begin{eqnarray}
\label{H12}
&&H_{1} (y)=1-q  \nonumber\\
&&+{qy\over\av{j}}\sum_{k}( BAj(k) + (1-A)\av{j} )P_{out}(k)H_{1} (y)^k\;,
\end{eqnarray}
and
\begin{equation}
\label{H02}
H_{0}(y)=1-q+qy\sum_{k}P_{out}(k)H_{1} (y)^k\;.
\end {equation}
If $A\rightarrow0$, one expects that $H_{0}(y)=H_{1}(y)$, since there is no
correlation between $j$ and $k$, thus the probability to have $k$
outgoing edges is $P_{out}(k)$ whether we choose the site randomly or
weighted by the incoming edges $j$.

$H_0(1)$ is the probability to reach an outgoing component of any {\it
finite\/} size choosing a site.  Thus, below the percolation
transition $H_{0}(1)=1$, while
above the transition there is a finite probability to follow a
directed link to a site which is a root of an
infinite outgoing component: $P_\infty=1-H_0(1)$. It follows that
\begin{equation}
\label {p_inf_perc}
P_\infty (q)=q(1-\sum_{k}^{\infty} P_{out}(k) u^k)\;,
\end {equation}
where $u\equiv H_1(1)$ is the smallest positive root of
\begin{eqnarray}
\label{u_perc}
&&u=1-q \nonumber\\
&&+{q\over\av{j}}\sum_{k}( BAj(k) + (1-A)\av{j} )P_{out}(k)u^k\;.
\end{eqnarray}
Here $P_\infty (q)$ is the fraction of sites from which an infinite
number of sites  is reachable.
Eq.~(\ref{u_perc})  can be solved numerically and the solution may be
substituted into Eq.~(\ref{p_inf_perc}), yielding the size of the
IN component at dilution $p=1-q$.

Near criticality, the probability to start from a site and reach a giant
outgoing component follows
$P_\infty\sim(q-\qc)^\beta$. For mean-field systems (such as
infinite-dimensional systems, random graphs and Cayley trees) 
it is known that $\beta=1$~\cite{frojdh}.
This regular mean-field result is not always valid. Instead,
following~\cite{cohencond} we study the behavior of Eq.~(\ref{u_perc}) near
$q=\qc$, $u=1$, and find
\begin{equation}
\label{beta}
\beta=\cases{
{1\over 3-\l^{\star}} &${2<\l^{\star}<3},$\cr
{1\over\l^{\star}-3} &${3<\l^{\star}<4},$\cr
1            &${\l^{\star}>4}, $}
\end{equation}
where
\begin{equation}
\label{boundary2}
\l^\star=\l_{out}+{\l_{in}-\l_{out}\over\l_{in}-1}\;.
\end{equation}
We see that the order parameter exponent $\beta$ attains its usual
mean-field value only for $\l^{\star}>4$. As $\l_{out}\rightarrow\l_{in}$ the
correlated fraction $BA$ of sites resembles non-directed
networks~\cite{cohencond,Vespignani} (where there is no distinction between 
incoming and outgoing degrees).  In this case  we
get $\l^\star=\l_{out}=\l_{in}$
for any amount of correlation $A$.
The criterion for the existence of a giant component is then
$\av{k^2}/\av{k}=1$, and not $2$ as in the non-directed case. The difference
stems from the fact that in the non-directed case one of the links is used to
reach the site, while in the directed case there is generally no correlation
between the location of the incoming and outgoing links. Therefore, one
more outgoing link is available for leaving the site.

Without any correlations, $A=0$, different terms prevail
in the analysis and
\begin{equation}
\label{betanocorre}
\beta=\cases{
{1\over\l_{out}-2} &${2<\l_{out}<3},$\cr
1            &${\l_{out}>3}. $}
\end{equation}
This is the same as Eq.~(\ref{beta}) but with $\l^{\star}=\l_{out}+1$.

The GSCC is the intersection of the IN and OUT components. Therefore, it
behaves as the smaller of the two components:
$\beta_{GSCC}=max(\beta_{in},\beta_{out})$. This can be also derived by 
applying the same methods as for the IN and OUT components to the generating 
function of 
the GSCC obtained in~\cite{dor_samu}. The exponent for the 
GWCC, on the other hand, is independent of the exponents of the other 
components, since the transition point is different.

It is known that for a random graph of
arbitrary  degree distribution the finite clusters
follow the scaling form
\begin {equation}
n(s)\sim s^{-\tau}e^{-s/s^*}\;,
\end{equation}
where $s$ is the cluster size and $n(s)$ is the number of clusters of
size $s$. At criticality $s^*\sim|q-\qc|^{-\sigma}$ diverges and the tail
of the  distribution follows  a power law.

The probability that $s$ sites can be reached from a site by following links 
at criticality follows $p(s) \sim s^{-\tau}$, and is generated by $H_0$, where
$H_0(y)=\sum_s p(s) y^s$.
As in ~\cite{cohencond}, $H_0(y)$ can be expanded from Eq.~(\ref{H01}).
In the presence of correlations we find
\begin{equation}
\tau=\cases{
{1+{1\over\l^{\star}-2}}
&${2<\l^{\star}<4},$\cr \frac{3}{2} &${\l^{\star}>4}.$ }
\end{equation}
The regular mean-field exponents are recovered for $\l^{\star}>4$.
For the uncorrelated case we get
\begin{equation}
\tau=\cases{
{1+{1\over\l_{out}-1}}
&${2<\l_{out}<3},$\cr \frac{3}{2} &${\l_{out}>3}.$ }
\end{equation}
Now the regular mean-field results are obtained for $\l>3$.

In summary, we calculate the percolation properties of directed scale-free
networks. We find that the percolation critical exponents in
scale-free networks are strongly dependent upon the existence of correlations
and upon the degree distribution exponents in the range of
$2<\lambda^{\star}<4$. This regime characterizes most naturally occurring
networks, such as metabolic networks or the WWW.
The regular mean-field behavior of
percolation in infinite dimensions is recovered only for $\l^{\star}>4$.
A connection is found between non-directed and directed scale-free
percolation exponents for any finite correlation between the {\it in}- and
{\it out}-degrees. In the uncorrelated case,
i.e. $P(j,k)=P_{in}(j)P_{out}(k)$, the probability to reach an
outgoing component does not bear any dependence upon $P_{in}(j)$. The results
are summarized in Table \ref{tb1}.

\begin{table}
\begin{tabular}{|c|c|c|} %\hline
& \emph{uncorrelated} & \emph{correlated}\\ \hline
\emph{GWCC} & $min(\l_{out},\l_{in})+1$ & $min(\l_{out},\l_{in})$\\
\emph{IN} & $\l_{out}+1$ & $\l_{out}+{\frac{\l_{in}-\l_{out}}{\l_{in}-1}}$\\
\emph{OUT} &  $\l_{in}+1$ & $\l_{in}+{\frac{\l_{out}-\l_{in}}{\l_{out}-1}}$\\
\emph{GSCC} & $min(\l_{out},\l_{in})+1$ & $min(\l^*_{out},\l^*_{in})$\\
%\hline
\end{tabular}
\narrowtext
\caption{Values of $\l^{\star}$ for the different network components for both
correlated and uncorrelated cases.}
\label{tb1}
\end{table}

\acknowledgements
Support from the NSF is gratefully acknowledged (DbA).

\end{multicols}
\end{document}